\catcode`\@=11

\newif\if@fewtab\@fewtabtrue
{\count255=\time\divide\count255 by 60
\xdef\hourmin{\number\count255}
\multiply\count255 by-60\advance\count255 by\time
\xdef\hourmin{\hourmin:\ifnum\count255<10 0\fi\the\count255}}
\def\ps@draft{\let\@mkboth\@gobbletwo
    \def\@oddhead{}
    \def\@oddfoot{\hbox to 7 cm{\tiny \versionno
       \hfil}\hskip -7cm\hfil\rm\thepage \hfil {\tiny\draftdate}}
    \def\@evenhead{}\let\@evenfoot\@oddfoot}

\def\draftcite#1{\ifnum\draftcontrol=1#1\else{}\fi}
\def\@lbibitem[#1]#2{\item{}\hskip -3cm \hbox to 2cm
{\hfil$\scriptstyle\draftcite{#2}$}\hskip
1cm[\@biblabel{#1}]\if@filesw
     {\def\protect##1{\string ##1\space}\immediate
      \write\@auxout{\string\bibcite{#2}{#1}}}\fi\ignorespaces}

\def\@bibitem#1{\item\hskip -3cm \hbox to 2cm
{\hfil {\footnotesize\draftcite{#1}}}\hskip 1cm
\if@filesw \immediate\write\@auxout
       {\string\bibcite{#1}{\the\value{\@listctr}}}\fi\ignorespaces}

\def\citen#1{\if@filesw \immediate\write \@auxout
{\string\citation{#1}}\fi%
\@tempcntb\m@ne \let\@h@ld\relax \def\@citea{}%
\@for \@citeb:=#1\do {\@ifundefined {b@\@citeb}%
    {\@h@ld\@citea\@tempcntb\m@ne{\bf ?}%
    \@warning {Citation `\@citeb ' on page \thepage \space undefined}}%
    {\@tempcnta\@tempcntb \advance\@tempcnta\@ne
    \setbox\z@\hbox\bgroup\ifcat0\csname b@\@citeb \endcsname \relax
    \egroup \@tempcntb\number\csname b@\@citeb \endcsname \relax
    \else \egroup \@tempcntb\m@ne \fi \ifnum\@tempcnta=\@tempcntb
    \ifx\@h@ld\relax \edef \@h@ld{\@citea\csname b@\@citeb\endcsname}%
    \else \edef\@h@ld{\hbox{--}\penalty\@highpenalty
    \csname b@\@citeb\endcsname}\fi
    \else \@h@ld\@citea\csname b@\@citeb \endcsname \let\@h@ld\relax
\fi}%
\def\@citea{,\penalty\@highpenalty\hskip.13em plus.13em
minus.13em}}\@h@ld}
\def\@citex[#1]#2{\@cite{\citen{#2}}{#1}}%
\def\@cite#1#2{\leavevmode\unskip\ifnum
\lastpenalty=\z@\penalty\@highpenalty\fi%
  \ [{\multiply\@highpenalty 3 #1%
  \if@tempswa,\penalty\@highpenalty\ #2\fi}]}   %
\makeatother 

\def\be          {\begin{equation}}
\def\calm        {{\cal M}}

\def\calmz       {{\cal M}^{(1)}}
\def\cft         {conformal field theory}
\def\csa         {Cartan subalgebra}
\def\complex     {{\bf C}}
\def\ee          {\end{equation}}
\def\eE          {{\rm e}}
\newcommand\erf[1] {(\ref{#1})}
\def\g            {{\bf g}}
\def\ga          {g_a}
\def\gb          {g_b}

\def\ha          {h_a}
\def\hb          {h_b}
\newcommand\ide[2] {{\cal E}_{#1,#2}}
\def\ii          {{\rm i}}
\long\def\labl#1   {\label{#1}\ee \ifnum\draftcontrol=1
                   \mbox{ }\\[-12 mm]\query{#1}\\[5 mm] \fi}
\def\lie         {Lie algebra}
\def\nontriv     {non-trivial}
\long\def\query#1{\hskip 0pt{\vadjust{\everypar={}\small\vtop to
0pt{\hbox{}%
     \vskip -13pt\rlap{\hbox to
49.0pc{\hfil{\vtop{\hsize=8pc\tolerance=6000%
     \hfuzz=.5pc\rightskip=0pt plus 3em\noindent#1}}}}\vss}}}}%
\def\rank        {{\rm rank}}
\def\reals         {\mbox{${\sf I}\!{\sf R}$}}
\newcommand\sect[1]{\section{#1} \setcounter{equation}{0}}
\def\rep         {representation}
\def\Spin        {{\rm Spin}}
\def\spin        {{\rm spin}}
\newcommand\version[1] {\ifnum\draftcontrol=1
\typeout{}\typeout{#1}\typeout{}
                   \vskip3mm \centerline{\fbox{{\tt DRAFT -- #1 -- }
                   {\small\draftdate}}}
                   \vskip3mm \fi}
\def\Vee           {{\scriptscriptstyle\vee}}
\def\zet         {{\bf Z}}

\def\draftdate{\number\month/\number\day/\number\year\ \ \ \hourmin }

\global\def\draftcontrol{0}
\catcode`\@=12

\documentclass[12pt]{article}
\usepackage{amssymb,amsfonts}\usepackage{epsf}

  \hoffset
-22mm \topmargin= -13mm 

\begin{document}


\begin{flushright}  {~} \\[-15 mm]  {\sf hep-th/9711104} \\[1mm]
{\sf CERN-TH/97-326} \\[0 mm]
{\sf November 1997} \end{flushright}

\begin{center} \vskip 13mm
{\Large\bf A NOTE ON THE GEOMETRY}\\[2.8mm]
{\Large\bf OF CHL HETEROTIC STRINGS}\\[16mm]

{\large Wolfgang Lerche, Christoph Schweigert} \\[3mm]
CERN \\[.6mm] CH -- 1211~~Gen\`eve 23 \\[11mm]
{\large Ruben Minasian}  \\[3mm]
Department of Physics\\[.6mm] Yale University, New Haven, CT
06520\\[11mm]
{\large Stefan Theisen}  \\[3mm]
Sektion Physik, Universit\"at M\"unchen \\[.6mm] Theresienstra\ss e 37,
\
D -- 80333~~M\"unchen\\[27mm]
{{\bf Abstract}}
\end{center}

\begin{quote} We present a few remarks on disconnected components of
the moduli space of heterotic string compactifications on $T_2$. We
show in particular how the eight dimensional CHL heterotic string can
be understood in terms of topologically non-trivial $E_8\times E_8$ and
$\Spin(32)/\zet_2$ vector bundles over the torus, and that the
respective moduli spaces coincide. \end{quote}

\vfill
\begin{flushleft}  {~} \\[15 mm]  
{\sf CERN-TH/97-326} \\[0 mm]
{\sf November 1997} \end{flushleft}

\pagebreak

\sect{Introduction}

The recent developments in string theory have made clear that moduli
spaces of string compactifications are often connected. This is
particularly pronounced for theories with extended supersymmetry.
However, even for theories with extended supersymmetry it is in general
not true that {\em all} string compactifications to the same number of
dimensions and with the same amount of supersymmetry are continuously
connected. It is indeed well-known that there are disconnected
components of theories with $N=2$ supersymmetry in eight and six
dimensions, or $N=4$ supersymmetry in four
dimensions~\cite{CHL,chpo,otherdims}. It would certainly be interesting
to have a deeper geometrical understanding of the appearance of such
disconnected components, which typically lead to non-simply laced gauge
groups.

For compactifications of the heterotic string, disconnected components
arise because such compactifications require the choice of a
manifold $X$ together with a holomorphic vector bundle on $X$ with
structure group, given by either $G=E_8\times E_8$ or
$\Spin(32)/\zet_2$.  The moduli space of holomorphic vector bundles on
$X$ can in general have several disconnected components. This applies
in particular to the two gauge groups just mentioned when $X$ is a
torus. Indeed, for $T_2$ compactifications of the heterotic string to
$D=8$ there are two known components: the usual component, ${\cal
M}_{18,2}$, corresponds to the standard Narain moduli space based on
the lattice $\Gamma_{18,2}$, while the other component ${\cal
M}_{10,2}$ corresponds to the CHL string whose moduli space is based on
$\Gamma_{10,2}$ \cite{CHL,chpo}. Moduli spaces of flat $G$-valued
connections over tori had also played a r\^ole in the recent work of
Friedman,  Morgan and Witten~\cite{FMW}. Quite generally, disconnected
components of moduli spaces correspond to topologically non-trivial
$G$-bundles.\footnote{As usual in physics, we call these bundles, by
slight abuse of language, topologically non-trivial, because they
cannot be deformed as {\em flat} bundles to the trivial bundle. This
should not be seen as a statement about   the topology of these
bundles.} Fortunately, a rather explicit description of all such
disconnected components is known from \cft \cite{schW3}. This will be
reviewed and further discussed in Section 2 of the present letter.

Subsequently, in section 3 we will then use this description to gain
geometrical insight into the nature of the CHL string in eight
dimensions.  In particular, we will explicitly show that the
disconnected components of the $E_8\times E_8$ and of the
$\Spin(32)/\zet_2$ compactification are equivalent, ie., that they form
the same moduli space ${\cal M}_{10,2}$. This generalizes the
well-known result \cite{gins} in the trivial component of the moduli
space, namely that ordinary toroidal compactifications of the two
heterotic strings map into the same moduli space ${\cal M}_{18,2}$.


\sect{Flat bundles with non-simply connected structure groups}

We start our discussion of disconnected components of the moduli space
with an explicit description of the moduli spaces of flat connections
over an elliptic curve $\Sigma$. We fix a compact, real, connected Lie
group $G$ which, however, is not necessarily simply connected. Its
universal covering group will be denoted by $\tilde G$; if
$Z:=\pi_1(G)$ is the fundamental group of $G$, then $G$ can be obtained
from $\tilde G$ by dividing out a subgroup of the center of $\tilde G$
that is isomorphic to $Z$:
\be G = \tilde G \, / \, Z \, . \ee

Let us fix a canonical basis $c_a,c_b$ of one-cycles on $\Sigma$. Flat
connections can be characterized by their holonomies $g_a$ and $g_b$
around $c_a$ and $c_b$ which are elements in the universal covering
group $\tilde G$. After projection to $G$ these elements have to
reproduce the single non-trivial relation in the fundamental group of
$\Sigma$:
\be
\pi(g_a g_b (g_a)^{-1} (g_b)^{-1}) = {\rm e}
\ee
which means in
$\tilde G$ that
\be
g_a g_b (g_a)^{-1} (g_b)^{-1} =
\omega\quad\mbox{with}\quad \omega\in Z \, .
\ee
As a consequence, the
moduli space $\calm_G$ of flat $G$-connections over $\Sigma$ decomposes
into $|Z|$ disconnected components, which are labelled by elements of
$Z$:
\be
\calm_G = \dot\cup  \calm_G^\omega \, .
\ee

It has been shown in \cite{schW3} that the components with $\omega\neq
{\rm e}$ -- which, as explained above, we call topologically
non-trivial components -- are isomorphic (as varieties, and up to a
rescaling also as complex spaces) to the topologically {\em trivial}
component of the moduli space over the same elliptic curve, but with
{\em another} structure group $G^\omega$. In other words: we have a
fully non-perturbative identity which allows us to trade topological
non-triviality for a change in the gauge group.

The change in the structure group corresponds to folding the {\em
affine} Dynkin diagram\footnote{This is related to but different as
ref.~\cite{FMW}, where automorphisms of non-affine Dynkin diagrams were
considered.}  by the automorphism associated to $\omega\in Z$. The
relevant folding data of all the groups are listed in the following
table:

  \be   \begin{tabular}{|l|c|c|lll|}
  \hline &&&&&\\[-.9em]
  \multicolumn{1}{|c|} {$G$} &
  \multicolumn{1}{c|}  {$\omega$} &
  \multicolumn{1}{c|}  {$N$} &
  \multicolumn{3}{c|}  {$G^\omega$}
  \\[1.2mm] \hline\hline &&&&&\\[-2.8mm]
   $A_n$      & $\!(\sigma_{n+1})^{(n+1)/N}\!$ & $N\!<\!n\!+\!1$
                   & \multicolumn{3}{c|} {$A_{((n+1)/N)-1}$}
\\[1.9mm]
   $A_n$      & $\sigma_{n+1}$ &$n+1$&&&$ \{ 0 \}$  \\[1.9mm]
   $B_{n}$    & $\sigma_v$    & 2  &&& $ \tilde B_{n-1}^{(2)}$
\\[1.9mm]
   $C_{2n}$   & $\sigma$  & 2  &&& $ \tilde B_{n}^{(2)} $ \\[1.9mm]
   $C_{2n+1}$ & $\sigma$  & 2  &&& $ C_n $             \\[1.9mm]
   $D_n$      & $\sigma_v$    & 2  &&& $ C_{n-2}$          \\[1.9mm]
   $D_{2n}$   & $\sigma_s$    & 2  &&& $ B_n $             \\[1.9mm]
   $D_{2n+1}$ & $\sigma_s$    & 4  &&& $ C_{n-1}$          \\[1.9mm]
   $E_6$      & $\sigma$  & 3  &&& $ G_2 $             \\[1.9mm]
   $E_7$      & $\sigma$  & 2  &&& $ F_4 $            
   \\[.4em] \hline \end{tabular} \ee

Two explicit examples are shown in Fig.1: in the left column it is
shown that the two non-trivial elements in the center of $E_6$ give the
moduli space of flat $G_2$ connections. The total moduli space of
topologically trivial and non-trivial $E_6$ connections thus has the
following three components:
\be
\calm_{E_6} =
\bigcup_{i=-1,0,1}\calm_{E_6}^i\ \cong\ \calm_{E_6}^0 \cup
\calm_{G_2}^0
\cup \calm_{G_2}^0  \, .
\ee

\goodbreak

Two comments are in order: we first remark that the case of $A_n$
corresponds to vector bundles of rank $n+1$. It seems that to algebraic
geometers the following closely related isomorphism has been known: the
moduli space $\calm(r,d)$ of vector bundles of rank $r$ and degree $d$
over an elliptic curve is isomorphic to the moduli space
$\calm(x(r,d),0)$, where $x(r,d)$ denotes the greatest common divisor
of $r$ and $d$. Also notice (for $A_n$) that if $\omega$ generates the
whole center, then the moduli space is just a single point. Our second
comment is about the symbol  $\tilde B_{n}^{(2)}$ (c.f., right part of
Fig.1): this denotes the only twisted affine \lie\ which has characters
that span a unitary \rep\ of the modular group. In this case, the
moduli space can be build from the maximal torus of this algebra and
its affine Weyl group precisely in the same way as the other moduli
spaces can be built from the data of the untwisted affine \lie s.

\begin{figure}[tbh]
\epsfysize=5cm \begin{center} \leavevmode
\epsfbox{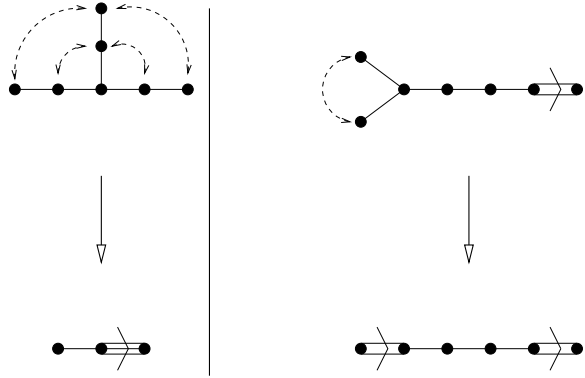}
\caption{Some foldings of affine Dynkin diagrams.}
\end{center} \end{figure}

More specifically, any connection in the topologically trivial sector
can be described by two commuting elements $\ga,\gb$ which are only
determined up to a simultaneous conjugation with an element of $G$;
hence we can assume without loss of generality that they are both
elements of a fixed maximal torus of $G$. We write
\be
\ga =
\exp(\ii\ha) \quad\mbox{and}\quad\gb=\exp(\ii\hb) \, .
\ee
The remaining gauge transformations are taken into account by the
diagonal action of the Weyl group $W$ on both maximal tori, so that the
moduli space is
\be
\calm_G = (\g_0 / L \oplus \g_0 / L ) / W
\ee
where $\g_0$ is the \csa\ of the \lie\ of $G$, and $L$ is the
coroot-lattice so that $\g_0/L$ is just the maximal torus. In the case
of $\tilde B^{(2)}_r$, $\g_0$ is an abelian \lie\ of rank $r$, $L$ is
the root lattice (not the {\em co}root lattice) of $B_r$, and $W$ is
the Weyl group of $\Spin(2r+1)$ or, equivalently, of ${\rm Sp}(2r)$. It
was shown in \cite{schW3} that the disconnected components of the
moduli space can be described using the analogous data of the group
$G^\omega$.

Let us make this explicit for the case of $\Spin(32)$, which is
relevant for the eight dimensional heterotic string compactifications
to be discussed later. Usually one describes the \lie\ of $\Spin(32)$
by antisymmetric matrices; here we prefer to work in a basis where the
Cartan subalgebra is realized by diagonal matrices. This is achieved by
the introduction of complex fermions $(\psi_i,\bar\psi_i), i=1\ldots
16$. In this basis, the boundary conditions for a world sheet of genus
$1$ and modular parameter $\tau$ and for winding numbers $L_1$ and
$L_2$ in the bosonic sector are:
\be
\psi_i (z+n+m\tau) = \exp(2\pi\ii (n\ha^i L_1 + m\hb^i
L_2)) \psi_i(z) \, ,
\ee
The corresponding partition function can be
computed explicitly \cite{nasw}.

We now extend this calculation to the topologically \nontriv\ bundles.
They are again described by the monodromies $\ga$ and $\gb$ along the
\nontriv\ cycles of the torus.  Here $\ga$ and $\gb$ are group elements
in $\Spin(32)$ which obey the condition \be \ga \gb \ga^{-1} \gb^{-1} =
\omega\, , \labl{desc} where $\omega$ is the element in the center of
$\Spin(32)$ that corresponds to the spinor conjugacy class. Notice that
in contrast to the topologically trivial sector, the monodromies now do
not commute any more, and the relations of the fundamental group
$\pi_1(T^2) \cong \zet\times\zet$ are only reproduced after projection
to $\Spin(32)/\zet_2$. According to the automorphism presented above,
$\calm_1$ is isomorphic to the moduli space of  topologically {\em
trivial} $\Spin(17)$-bundles. Hence its dimension is $\dim_\complex
\calmz = \rank (\Spin(17)) = 8$.

More generally, a simultaneous conjugation of $\ga$ and $\gb$ with an
element in $\Spin(2n)$ can be used to bring the general solution of
\erf{desc} to the form \be \ga= \exp(\ii (h_0 + h_a) \quad \gb = A
\exp(\ii h_b) \ee where $h_0$, $\ha$ and $\hb$ are elements in the
\csa\ of $\spin(2n)$ and $A$ is a group element which upon conjugation
reproduces on the \csa\ the action of the following Weyl
transformation:
\be \begin{array}{lll}
w(\alpha^{(i)}) &=& \alpha^{(n-i)} \quad\mbox{for}\quad i=1\ldots n-1
\\
w(\alpha^{(n)}) &=& -\theta = - \alpha^{(1)} -\alpha^{(n-1)}
-\alpha^{(n)}
 -2 \sum_{i=2}^{n-2} \alpha^{(i)}
\, . \end{array}\labl{weyl}
$h_0$ is a specific element of the \csa\ which is described in more
detail in \cite{schW3}. Moreover, the elements $\ha$ and $\hb$ are
restricted by the condition
\be
A h_a A^{-1} = h_a \quad\mbox{and}\quad
A h_b A^{-1} = h_b \, .
\ee
This condition implies that only half of the possible Wilson lines in
the topologically trivial sector survives. This accounts for the
reduction of the dimension of the moduli space.

An explicit description of the possible Wilson lines $h_a$ and $h_b$ is
as follows (for notation, details and a derivation see the appendix):
in a suitable basis the general element of the \csa\ can be written as
\be \sum_{i=1}^n \lambda_i(\ide ii - \ide{i+n}{i+n}) \, . \ee where
$\ide_{ij}$ is the $2n\times2n$ matrix with only zero entries,
except 
one. The subalgebra of the \csa\ that is invariant under conjugation
with $A$ is given by those elements for which \be \lambda_i =
\lambda_{n+1-i}  \, ;\ee it is an $n/2$-dimensional subalgebra.

\sect{Geometric realization of the 8d CHL compactification}

We now come to compactifications of the heterotic string to 8
dimensions. The compactification space has to be a Calabi-Yau manifold,
and hence be the two-dimensional torus. Moreover, one has to choose a
gauge bundle over this compactification manifold. The conditions for
anomaly cancellation require the structure group of this bundle to be
either $G_1=E_8\times E_8$ or $G_2=\Spin(32)/\zet_2$. Since the
Calabi-Yau metric on the torus is flat, anomaly cancellation tells us
that the background gauge field on the bundle has to be flat as well.

Let us consider first the case of $G_2=\Spin(32)/\zet_2$. This group is
not simply connected, and we have seen that in this case the moduli
space $\calm$ of flat connections has more than one connected
component. More precisely, one finds that
$\pi_0(\calm)=\pi_1(G_2)=\zet_2$. A similar logic applies to
$G_1=E_8\times E_8$; in this case the outer automorphism is just the
permutation of the two $E_8$ factors, and again we have two connected
components of the moduli space.

We will now show that the component $\calm_1$ gives the CHL
compactification of the heterotic string. This affords in particular a
geometric description of the CHL compactification.

To fix the Wilson lines, we choose $n/2$ elements $\theta_i\in
\reals\bmod\zet$ and $n/2$ elements $\phi_i\in \reals\bmod\zet$.
Given a world sheet of genus $1$ and modular parameter $\tau$ and in
the sector with winding numbers $L_1$ and $L_2$ in the bosonic sector,
the boundary conditions for the  $2n$ complex fermions are
in the direction of the $a$-cycle
\be
\psi^i(z+1) = R(\eE^{2\pi\ii L_1 h_a})^{ii}\psi^i(z) = \left\{
\begin{array}{ll}
\eE^{2\pi\ii\theta_i L_1}\psi^i(z) &\quad\mbox{for}\quad i=1\ldots n/2
\\
\eE^{-2\pi\ii\theta_{n+1-i}L_1}\psi^i(z) &\quad\mbox{for}\quad i=n/2+1
\ldots n \\
\eE^{-2\pi\ii\theta_{i-n}L_1}\psi^i(z) &\quad\mbox{for}\quad i=n+1
\ldots 3/2 n \\
\eE^{2\pi\ii\theta_{2n+1-i}L_1}\psi^i(z) &\quad\mbox{for}\quad i=3/2n+1
\ldots 2 n
\end{array}\right. \ee
and in the direction of the $b$-cycle for odd winding $L_2$
\be
\psi^i(z+\tau) = \sum_{j=1}^{2n} R(A^{L_2}\eE^{2\pi\ii h_b L_2})^{ij}
\psi_j(z)=
\left\{ \begin{array}{ll}
\ii\eE^{2\pi\ii\phi_iL_2}\psi^{2n-i}(z) &\quad\mbox{for}\quad i=1\ldots
n/2 \\
\ii\eE^{-2\pi\ii\phi_{n+1-i}L_2}\psi^{2n-i}(z) &\quad\mbox{for}\quad
i=n/2+1 \ldots n \\
-\ii\eE^{-2\pi\ii\phi_{i-n}L_2}\psi^{2n-i}(z)
&\quad\mbox{for}\quad i=n+1 \ldots 3/2 n \\
-\ii\eE^{2\pi\ii\phi_{2n+1-i}L_2}\psi^{2n-i}(z) &\quad\mbox{for}\quad
i=3/2n+1
\ldots 2 n
\end{array}\right. \ee
and similarly for even winding.
We see that the fermions come in groups of four, and the boundary
conditions of such a group of four fermions is described by two
parameters $\theta_i$ and $\phi_i$. For $\Spin(32)/\zet_2$ we have the
following groups of indices: $(1,16,8,9)$, $(2,15,7,10),$
$(3,14,6,11)$,
$(4,13,5,12)$.

This should be compared to the topologically non-trivial component of
the moduli space of the $E_8\times E_8$ string. Here, we impose the
following boundary conditions: along one cycle, say $c_a$, we introduce
an ordinary Wilson line, while the monodromy along the other cycle
$c_b$ interchanges the two $E_8$ factors and hence in particular the
fermionic operators that are used to make up their \csa s. This leads
to an exchange of two complex fermions, which in turn can be described
in terms of real fermions. Up to different phases (which can be
compensated by the Wilson lines and a relabeling of the fermions), we
recover exactly the same behaviour as for the non-trivial component of
the $\Spin(32)/\zet_2$ string.

We have thus shown that the CHL string has a natural geometric
description and that the equivalence of heterotic compactifications
based $E_8\times E_8$ on the one hand and on $\Spin(32)/\zet_2$ on the
other hand holds in the topologically non-trivial component of the
moduli space of connections as well.

Notice that so far we have been concerned only with the moduli space
$\calm_G$ of flat connections, i.e.\ the moduli space of the Wilson
lines. The full moduli space of heterotic compactifications is locally
a product of this moduli space and the moduli space $\calm_{2,2}$ of
2-tori:
\be
\begin{array}{lll}
\calm_{18,2} &\sim& \calm_{{\Spin(32)}} \times \calm_{2,2}\ \sim\
  \calm_{{ E_8\times E_8}} \times \calm_{2,2}
\\
\calm_{10,2} &\sim& \calm_{{\Spin(17)}} \times \calm_{2,2}/\zet_2
\, . \end{array}
\labl{global}
In this context, the following observation seems to be intriguing: for
the Narain compactification, the moduli space is described by either
$E_8\times E_8$ or $\Spin(32)$ bundles, and both groups appear as
possible gauge enhancements (by allowing general rotations involving
$\Gamma_{2,2}$, one can extend $\Spin(32)$ to $\Spin(36)$). In the CHL
compactification, the Wilson lines are described by $\Spin(17)$ which
is not simply-laced and does not appear as a possible gauge group of
the CHL-string. However, its dual (obtained by reversing the arrow in
the [non-extended] Dynkin diagram), which is ${\rm Sp}(16)$, does (by
allowing general rotations, this can similarly be extended to
$Sp(20)$). In other words, and this seems to be a general rule, the
gauge group that one canonically obtains is not given by the structure
group $G$ of the bundle, but by its dual, $G^\Vee$.

A similar structure, however related to affine Dynkin diagrams,
was found in ref.\ \cite{FMW}. In fact, this pattern
is familiar from many contexts, like for example in $N=2$
gauge theories \cite{MW}.

Given that the CHL string has this geometric interpretation
which puts it on the same footing as the usual Narain compactification,
it is natural to ask what its $F$-theory dual is; this will be
addressed in a future publication.

\vskip 2cm
\noindent {\bf Acknowledgements}\\
We would like to thank Peter Mayr for collaborating on $F$-theory
related aspects of this work.

\vskip 2cm\goodbreak


  \newcommand{\wb}{\,\linebreak[0]} \def\wB {$\,$\wb}
  \newcommand{\Bi}[1]    {\bibitem{#1}}
  \newcommand{\Erra}[3]  {\,[{\em ibid.}\ {#1} ({#2}) {#3}, {\em
Erratum}]}
  \newcommand{\BOOK}[4]  {{\em #1\/} ({#2}, {#3} {#4})}
  \newcommand{\vypf}[5]  {{#1} [FS{#2}] ({#3}) {#4}}
  \newcommand{\J}[5]   {{#1} {#2} ({#3}) {#4} }
  \newcommand{\Prep}[2]  {{\sl #2}, preprint {#1}}
 \def\phrd  {Phys.\wb Rev.\ D}
 \def\nupb  {Nucl.\wb Phys.\ B}
 \def\phlb  {Phys.\wb Lett.\ B}


\vskip 2cm\goodbreak

\appendix
\sect{Conjugations in $\Spin(2n)$}

We need the group elements $\ga$ and $\gb$ that describe the
monodromies around $c_a$ and $c_b$ explicitly in the $2n$-dimensional
representation of $\Spin(2n)$, the vector representation. Notice that
this is {\em not} a representation of $\Spin(2n)/\zet_2$, and as a
consequence, the element $\omega$ in the center is not represented by
the identity.

To this end we first give the generators of the Lie algebra $so(2n)$ in
the $2n$-dimensional representation. Denote by $\ide ij$ the $2n\times
2n$ matrix which has only zeros as entries, except for 1 in the $i$-th
row and $j$-th column. Again it is convenient not to work with
antisymmetric matrices but to perform a unitary transformation with the
matrix \be U:= \frac1{\sqrt2}\sum_{i=1}^n \ii \ide ii - \ide {i+n}{i+n}
-\ii\ide i{n+i} -\ide{n+i}i \, . \ee This corresponds to the usual
introduction of complex fermions and allows to represent elements of
the \csa\ by diagonal matrices. The matrices $B$ of the
$2n$-dimensional \rep\ of the Lie group $\Spin(2n)$ are then
characterized by the fact that $B^tK B =K$, where $K$ is the matrix \be
K:= U^t U = \sum_{i=1}^n  \ide i{i+n} + \ide {i+n}i \, . \ee

The generators of the Lie algebra $\spin(2n)$ in a Cartan-Weyl basis
are then
represented as follows:
\be \begin{array}{lll}
R(E^i_+) &=& \ide i{i+1} - \ide{i+n+1}{i+n} \quad\mbox{for}\quad
i=1\ldots n-1 \\
R(E^i_-) &=& \ide {i+1}i - \ide{i+n}{i+n+1} \quad\mbox{for} \quad
i=1\ldots n-1 \\
R(E^n_+) &=& \ide {n-1}{2n} - \ide{n}{2n-1} \\
R(E^n_-) &=& \ide {2n}{n-1} - \ide{2n-1}n \\
R(H^i) &=& \ide ii - \ide{i+1}{i+1} -\ide{i+n}{i+n}+\ide{i+n+1}{i+n+1}
\quad\mbox{for} \quad i=1\ldots n-1 \\
R(H^n) &=& \ide {n-1}{n-1} + \ide nn -\ide{2n-1}{2n-1}-\ide{2n}{2n}
\, . \end{array}\ee
Finally, we need the generators for the $su(2)$-subalgebra
of $\spin(2n)$ generated by the highest root $\theta$ of $\spin(2n)$:
\be \begin{array}{lll}
R(E^\theta_+) &=& \ide 1{n+2} - \ide{2}{n+1} \\
R(E^\theta_-) &=& \ide {n+2}{1} - \ide{n+1}2 \\
R(H^\theta) &=& \ide 11 + \ide 22 -\ide{n+1}{n+1}-\ide{n+2}{n+2}
\, . \end{array}\ee

We claim that the following matrix is a solution: \be A := \sum_{i=1}^n
-\ii \ide i{2n+1-i} + \ii \ide{2n+1-i}i \, . \ee One can check that it
has the following properties: $A^2=1$, $A$ is in $SO(2n)$, which
follows from the fact that \be U A U^+ = - \sum_{i=1}{2n} \ide
i{2n+1-i} \, \ee which is a real orthogonal matrix. Moreover, one has
\be \begin{array}{lll}
AR(E^i_\pm) A &=& - R(E^{n-i}_\pm) \quad\mbox{for}\quad i=1\ldots n-1
\\
AR(E^n_\pm) A &=&  R(E^{\mp\theta})  \\
AR(H^i) A &=& R(H^{n-i}) \quad\mbox{for}\quad i=1\ldots n-1 \\
AR(H^n) A &=&  R(H^{\theta})
\, . \end{array}\ee
This reproduces correctly the action of the Weyl group element $w$
described
in \erf{weyl}.
\end{document}